\documentclass[letterpaper,twocolumn,english,revtex]{revtex4-1}
\usepackage[T1]{fontenc}
\usepackage[latin9]{inputenc}
\usepackage{amsmath}
\usepackage{amssymb}
\usepackage{graphicx}

\makeatletter

\pdfpageheight\paperheight
\pdfpagewidth\paperwidth

\makeatother

\usepackage{babel}
\begin{document}
\title{\noindent Landauer's erasure principle in a squeezed thermal memory}
\author{\noindent Jan Klaers$^{1}$}
\email{j.klaers@utwente.nl}

\address{\noindent $^{1}$Complex Photonic Systems (COPS), MESA$^{+}$ Institute
for Nanotechnology, University of Twente, 7522 NB Enschede, The Netherlands}
\begin{abstract}
\noindent Landauer's erasure principle states that the irreversible
erasure of a one-bit memory, embedded in a thermal environment, is
accompanied with a work input of at least $k_{\text{B}}T\ln2$. Fundamental
to that principle is the assumption that the physical states representing
the two possible logical states are close to thermal equilibrium.
Here, we propose and theoretically analyze a minimalist mechanical
model of a one-bit memory operating with squeezed thermal states.
It is shown that the Landauer energy bound is exponentially lowered
with increasing squeezing factor. Squeezed thermal states, which may
naturally arise in digital electronic circuits operating in a pulse-driven
fashion, thus can be exploited to reduce the fundamental energy costs
of an erasure operation.
\end{abstract}
\maketitle
Energy dissipation is one of the main design considerations in digital
electronics today \citep{Frank2002,Haensch2006,Pop2010}. Smaller
transistors operating at lower-voltages are a natural design choice
that may reduce the power consumption of central processing units.
In 1961, Rolf Landauer argued that there exists a limit to which the
power consumption of certain logical operations can be reduced. Landauer's
principle states that the erasure (or reset) of one bit of classical
information is necessarily associated with an entropy increase of
at least $k_{\text{B}}\ln2$ and an energy input of at least $k_{\text{B}}T\ln2$
\citep{Landauer1961,Plenio2001,Vaccaro2011,Berut2012,Jun2014,Parrondo2015,Hong2016,Yan2018}.
For the present generation of silicon-based integrated circuits, the
energy dissipation per logic operation is about a factor of 1000 larger
than the Landauer limit. It is, however, predicted that the Landauer
limit will be reached within the next decades \citep{Frank2002,Haensch2006,Pop2010}.
Thus, improvements in our understanding of energy dissipation in information-processing
devices are of both scientific interest and of technological relevance.
Due to the ongoing miniaturization, non-equilibrium and quantum effects
must be taken into account \citep{Hilt2011,delRio2011,Esposito2011,Goold2015,Pezzutto2016,Millen2016,Manzano2017}.
In this work, it is theoretically demonstrated that memory devices
embedded in a squeezed thermal environment are unbounded by the Landauer
limit. In these environments, thermal fluctuations show fast periodic
amplitude modulations, which can be exploited to reduce the minimum
energy costs for an erasure operation below the standard Landauer
limit. This situation may naturally occur in digital electronic circuits
operating in a pulse-driven fashion and, in future, could be exploited
to build more energy-efficient electronic devices. 

Squeezed thermal states are the classical analog of squeezed coherent
states in quantum mechanics. Both states are characterized by an asymmetric
phase space density as opposed to the rotationally invariant phase
space densities of coherent, thermal, or vacuum states. A mechanical
oscillator may be prepared in a squeezed thermal state \citep{Fearn1988,Kim1989}
by a periodic modulation of the spring constant \citep{Rugar1991}.
This leads to a state with reduced thermal fluctuations in one quadrature
(e.g. momentum) and enhanced fluctuations in the orthogonal quadrature
(e.g. position). In the context of heat engines, squeezed thermal
reservoirs have been proposed as a resource for work generation unbounded
by the standard Carnot limit \citep{Rossnagel2014,Correa2014,Manzano2016,Niedenzu2016,Niedenzu2018}.
Due to the non-equilibrium nature of these reservoirs, this does not
violate the second law of thermodynamics. In recent work \citep{Klaers2017},
we have demonstrated a physical realization of such an engine, in
which the working medium consists of a vibrating nano-beam that is
driven by squeezed electronic noise to perform work beyond the Carnot
limit. 
\begin{figure*}
\begin{centering}
\includegraphics[width=17cm]{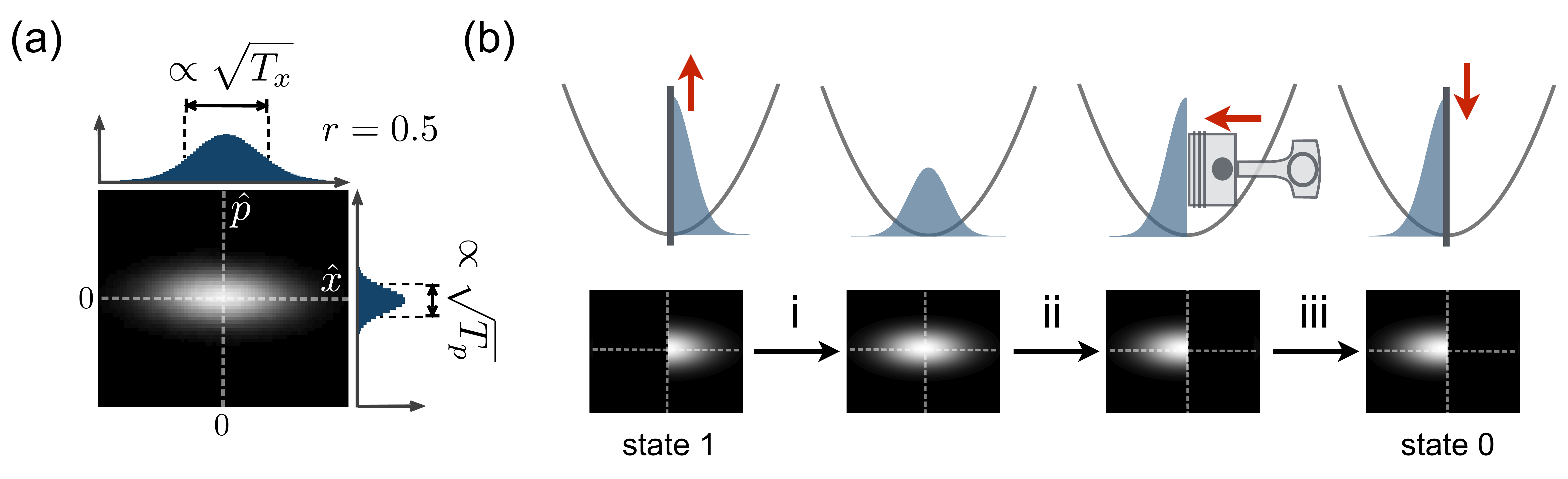}
\par\end{centering}
\caption{\label{fig:1}(a) Stroboscopic phase space probability density of
a squeezed thermal state (squeezing factor $r=0.5$, relative timing
$t_{0}=0.24\,\nu^{-1}$). The thermal fluctuations in one quadrature
are reduced, while the orthogonal quadrature shows increased fluctuations.
The variances of the line integrated distributions (shown in blue)
correspond to two temperatures $T_{x}$, $T_{p}$ that govern the
thermal fluctuations of the system. (b) Scheme to erase one bit of
information in a squeezed thermal memory: (i) removal of partition
and free expansion, (ii) stroboscopic compression to half of the volume
(iii) insertion of partition. During the process, the single-particle
gas is assumed to be in contact with the squeezed thermal reservoir
at all times.}
\end{figure*}
We have furthermore demonstrated that a phase-selective thermal coupling
allows to extract work from a single squeezed thermal reservoir, which
is not possible with a standard thermal reservoir \citep{Scully2001}. 

In this work, we propose and theoretically analyze a minimalist mechanical
model of a one-bit memory subject to squeezed thermal noise. This
memory consists of a single particle that is trapped in a harmonic
potential. The trap can be spatially divided into two halves by a
partition in the trap center. If the particle resides on the left-hand
side of the trap, the memory is regarded as being in the logical state
'0'; if it is located on the right hand side, the memory is in the
'1' state. We further assume that the particle is coupled to a squeezed
thermal reservoir, which can be modeled by introducing a stochastic
force $f=f(t)$ to its equation of motion, as described by the Langevin
equation $m\ddot{x}=F(x)-c\dot{x}+f$. Here $m$ denotes the mass
of the oscillator, $c$ is the viscous damping coefficient, and $F(x)=-m\omega_{0}^{2}x$
describes the restoring force with $\omega_{0}$ as the undamped oscillator
frequency. The stochastic force $f(t)$ is synthesized from two independent
noise signals $\xi_{1,2}(t)$ that are mixed with sine and cosine
component of a local oscillator at frequency $\omega=2\pi\nu$ \citep{Klaers2017}:
\begin{equation}
f(t)\hspace{-0.5mm}=\hspace{-0.5mm}a_{0}\hspace{-0.5mm}\left[\text{e}^{+r}\xi_{1}(t)\cos(\omega t)\hspace{-0.5mm}+\hspace{-0.5mm}\text{e}^{-r}\xi_{2}(t)\sin(\omega t)\right]\label{eq:stochastic_force}
\end{equation}
The squeezed thermal reservoir modeled by $f(t)$ is characterized
by an overall amplitude $a_{0}$ and a squeezing parameter $r$ that
tunes the imbalance between the two orthogonal quadratures. Assuming
$\xi_{1,2}(t)$ to be white noise, the power spectral density is frequency-independent
and increases exponentially with the squeezing factor: $\text{psd}[f](\omega)\propto\cosh2r$.
The squeezing introduces fast periodic amplitude modulations in the
stochastic force as can be seen from 
\begin{equation}
\left\langle f^{2}(t)\right\rangle =\frac{a_{0}^{2}}{3}\left[\text{e}^{-2r}+2\sinh(2r)\cos^{2}(\omega t)\right]\,\text{,}\label{eq:amplitude_modulation}
\end{equation}
in which $\left\langle \ldots\right\rangle $ denotes the statistical
average over many independent realizations. 

The impact of the so defined stochastic force on the single-particle
gas can be investigated by means of phase space densities, such as
presented in Fig. \ref{fig:1}a. The numerical results shown here,
and in the rest of this work, have been obtained by integrating the
Langevin equation using the Runge-Kutta method (fourth order) with
constant time steps. The local oscillator in eq. (\ref{eq:stochastic_force})
is assumed to be resonant with the undamped oscillator ($\omega=2\pi\nu=\omega_{0}$)
and the noise functions $\xi_{1,2}(t)\in\text{[-1;1]}$ are sampled
from white noise generated by a (pseudo-)random number generator with
a high-frequency cutoff at $\nu/2$. The phase space density presented
in Fig. \ref{fig:1}a demonstrates a reduction in thermal fluctuations
in the squeezed quadrature and an increase in the anti-squeezed quadrature
(squeezing factor $r=0.5$). The quantities $\hat{x}$ and $\hat{p}$,
corresponding to the two axes of the phase space plots, may be regarded
as two orthogonal quadratures co-rotating with the driving force (rotating
frame). Another valid interpretation is to regard $\hat{x}=x\sqrt{m\omega/\hbar}$
and $\hat{p}=p/\sqrt{\hbar\omega m}$ as dimensionless instances of
the actual physical position $x$ and momentum $p$ (laboratory frame).
In this case, the diagram in Fig. \ref{fig:1}a represents a stroboscopic
phase space density measured at equidistant points in time $t_{0},\,\nu^{-1}+t_{0},\,2\nu^{-1}+t_{0},\,\ldots$,
where $t_{0}$ sets the relative timing of the observations with respect
to the local oscillator in the stochastic force. For the remainder
of this work, we restrict our presentation to the laboratory frame.
An important consequence is that any interaction with the system has
to be performed in a stroboscopic fashion. A spatial compression,
for example, needs to be divided into a sequence of smaller compression
steps that have to be executed with the desired timing $t_{0}$. A
concrete realization of the latter is moving the piston with the velocity
$v(t)=v_{\text{max}}\cos^{2n}(\omega(t-t_{0}))$, in which $n$ is
a large positive integer and the maximum velocity $v_{\text{max}}$
is kept sufficiently small.

Squeezed thermal states can be understood in terms of a generalized
Gibbs ensemble \citep{Manzano2018,Klaers2017}. The thermal fluctuations
of the two orthogonal quadratures $\hat{x}$ and $\hat{p}$ are controlled
by two different temperatures $T_{x}$ and $T_{p}$, which take the
role of state variables (see Fig. \ref{fig:1}a). The corresponding
stroboscopic phase space density follows 
\begin{equation}
\rho_{\text{sq}}(\hat{x},\hat{p})\,\propto\,\exp\left(-\frac{\hbar\omega\hat{x}^{2}}{2k_{\text{B}}T_{x}}-\frac{\hbar\omega\hat{p}^{2}}{2k_{\text{B}}T_{p}}\right)\;\text{.}\label{eq:squeezed_thermal}
\end{equation}
An effective system temperature $T$ may be defined as $T=\sqrt{T_{x}T_{p}}$.
A consequence of this definition is that an isothermal squeezing operation
($T=\text{const}$) does not increase the entropy of the state \citep{Klaers2017}.

\begin{figure*}
\begin{centering}
\hspace{20mm}\includegraphics[width=15cm]{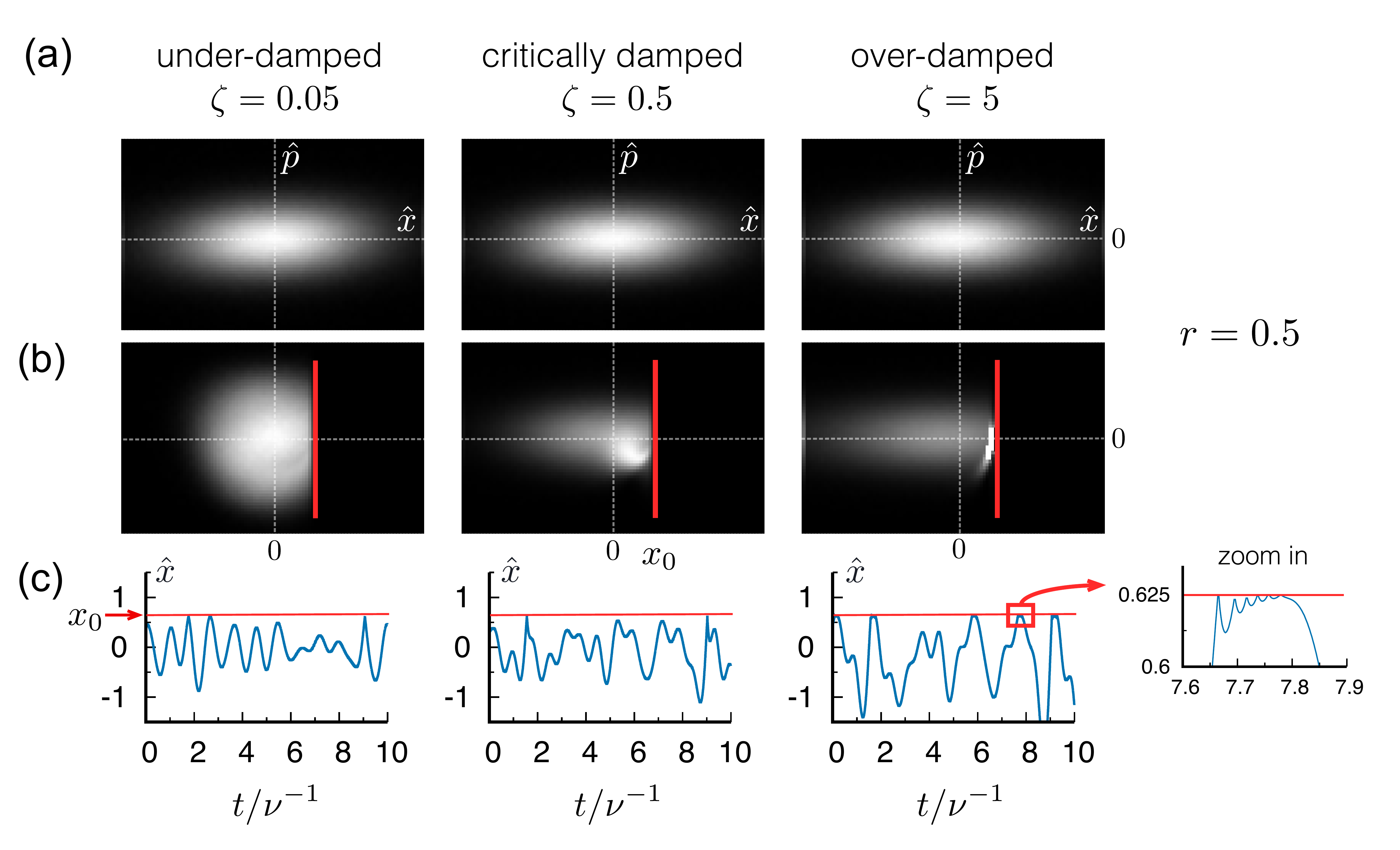}\vspace{-4mm}
\par\end{centering}
\caption{\label{fig:2}(a) Stroboscopic phase space densities of a harmonically
confined particle subject to squeezed thermal noise for three different
damping regimes: under-damped motion (damping ratio $\zeta=c/2m\omega_{0}=0.05$),
critically damped motion ($\zeta=0.5$), and over-damped motion ($\zeta=5$).
The diagrams represent position-momentum histograms at equidistant
points in time $t_{0},\,\nu^{-1}+t_{0},\,2\nu^{-1}+t_{0},\,\ldots$,
where $t_{0}=0.24\,\nu^{-1}$ sets the relative timing of the observations
with respect to the stochastic force. (b) Collisions with the piston
(indicated by the red bar) induce phase jumps in the particle motion,
which cancel the squeezing effect in the case of under-damped motion.
In the case of over-damped motion, the particle tends to 'stick' to
the piston, which leads to a strong enhancement of the probability
density in this region. (c) Typical trajectories $x(t)$ of the trapped
particle.}
\end{figure*}
The scheme to erase one bit of information in a squeezed thermal memory
is shown in Fig. \ref{fig:1}b. During the process, the single-particle
gas is assumed to be in contact with the squeezed thermal reservoir
at all times. First, the partition is removed and the gas expands
freely (step i in Fig. 1). In the second step (ii), the gas is compressed
by a piston. In the last step (iii), the partition is put back in
the center of the trap. This procedure initializes the memory in the
state 0 regardless of the initial conditions. The general idea behind
the scheme is to use squeezing as a means to reduce the occurrence
of large positive momenta at the position of the piston during the
compressions steps. The latter reduces the pressure and, thus, the
work required for the compression. In our analysis, we make several
assumptions: it is assumed that the removal (and insertion) of the
partition is free of any energy cost. The collisions of the particle
with the piston are considered fully elastic. We also assume that
the collisions leave the motional state of the piston essentially
unchanged. Note that the proposed scheme relies on the notion of a
spatially compressed squeezed thermal state. We will first discuss
some subtleties and apparent difficulties associated to the latter. 

Figures \ref{fig:2}a,b show numerically obtained phase space densities
of a confined single-particle gas subject to squeezed thermal noise
in the under-damped (damping ratio $\zeta=c/2m\omega_{0}=0.05$),
critically damped ($\zeta=0.5$), and over-damped regime ($\zeta=5$).
For purely harmonic confinement (Fig. \ref{fig:2}a), the response
of the gas to the squeezed noise is largely independent of the damping
regime. This is markedly different in the presence of a piston (Fig.
\ref{fig:2}b). Collisions of the particle with the piston induce
phase-shifts in the otherwise purely harmonic motion. In the under-damped
regime, these phase-shifts destroy the correlation between particle
motion and squeezed noise, which cancels the squeezing phenomenon.
In the critically damped and over-damped case, the collisions with
the piston perturb, but do not destroy the squeezing phenomenon. In
the over-damped region, an additional effect comes into play, namely,
that the particle tends to 'stick' to the piston, which leads to a
strong enhancement of the probability density in this region. This
effect can also be observed in Fig. \ref{fig:2}c, which shows typical
examples of the particle motion $x(t)$ in the various damping regimes.
In the over-damped case, the particle tends to collide several times
with the piston before it is finally accelerated in the opposite direction. 

\begin{figure}
\begin{centering}
\includegraphics[width=8.75cm]{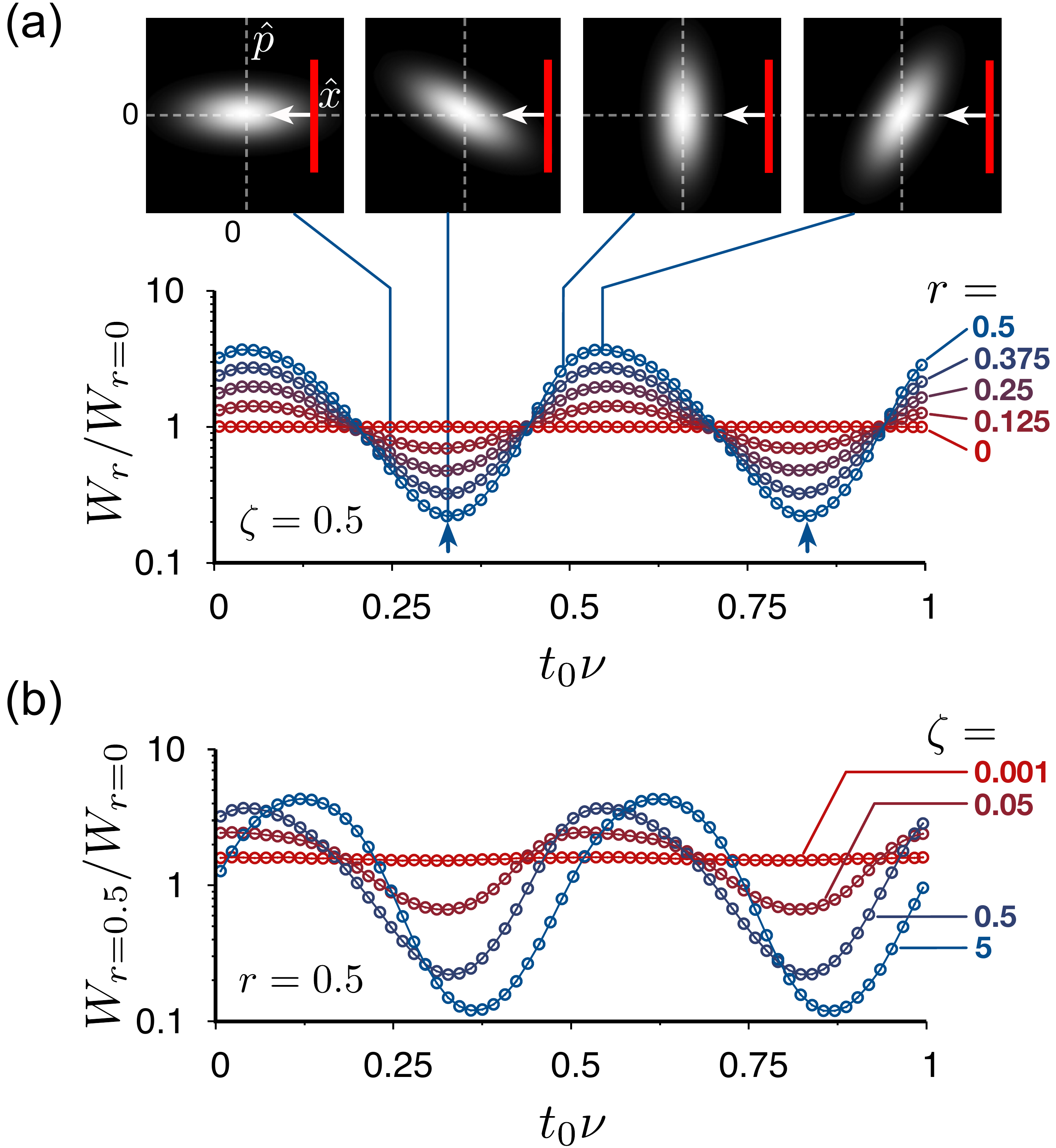}\vspace{-2mm}
\par\end{centering}
\caption{\label{fig:3}Required work $W$ to compress the single-particle gas
to half of its initial volume as a function of the relative timing
$t_{0}$, which defines the points in time, namely $t_{0},\,\nu^{-1}+t_{0},\,2\nu^{-1}+t_{0},\,\ldots$,
at which the compression steps are executed. The given values of $W_{r}$
(for different squeezing parameters $r$) are normalized to the work
at vanishing squeezing $W_{r=0}$. The temperature $T$ is kept fixed
in all simulations. (a) In the critically damped regime ($\zeta=0.5$),
the work is observed to exponentially decrease with the squeezing
parameter $r$ close to $t_{0}=0.33\,\nu^{-1}$ and $t_{0}=0.83\,\nu^{-1}$.
(b) Work at constant squeezing $r=0.5$ for various damping ratios
$\zeta$. The squeezing effect is observed to vanish for under-damped
systems.}
\end{figure}

By recording the elastic collision events in our numerical simulations,
we can derive the work $W$ required to compress the gas to half of
its initial volume. In Fig. \ref{fig:3}a, $W$ is shown as a function
of the parameter $t_{0}$, which defines the points in time, namely
$t_{0},\,\nu^{-1}+t_{0},\,2\nu^{-1}+t_{0},\,\ldots$, at which the
compression steps are executed.
At a relative timing around $t_{0}=0.33\,\nu^{-1}$ and $t_{0}=0.83\,\nu^{-1}$,
the work $W$ is found to exponentially decrease with the squeezing
parameter $r$ (note the logarithmic scale in Fig. \ref{fig:3}a).
Under those conditions, the squeezing effect reduces the occurrence
of large positive momenta close to the piston (indicated by the red
bar), which causes a reduced pressure exerted on the piston. Our numerical
results, thus, give clear evidence that squeezing can be exploited
to reduce the required work for the reset of a one-bit memory. Note
that this effect applies to both the critical and the over-damped
regime, but vanishes for strongly under-damped systems as shown in
Fig. \ref{fig:3}b.

In the remainder of this work, we discuss a simplifying analytical
model that captures the key aspects of the described phenomenon. The
presence of a piston at position $x_{0}$ introduces a cut-off in
phase space: $\rho(\hat{x}>x_{0},\hat{p})=0$. We will consequently
model a spatially compressed squeezed thermal state by the density
\begin{equation}
\rho(\hat{x},\hat{p})=Z^{-1}\,\rho_{\text{sq}}(\hat{x},\hat{p})\,\Theta(x_{0}-\hat{x})\;\text{,}\label{eq:phasespacedensity}
\end{equation}
in which $\Theta(x)$ is the Heaviside step function ($\Theta(x)=1$
for $x>0$, $\Theta(x)=0$ otherwise) and $Z$ is a normalization
constant such that $\iint\rho(\hat{x},\hat{p})\,d\hat{x}d\hat{p}=1$.
We choose to perform the compression step against a purely momentum
squeezed state of the gas as depicted in Fig. \ref{fig:2}a. To this
end, we set $T_{x}=T\exp(+2r)$ and $T_{p}=T\exp(-2r)$ \citep{Fearn1988}.
To derive the work for the compression, we start with a common ansatz
in the kinetic gas theory relating the pressure exerted on the piston
to the average momentum transfer by elastic collisions: $P=\int_{0}^{\infty}2\hbar\omega\hat{p}^{2}\rho(x_{0},\hat{p})\,d\hat{p}$.
Using eq. (\ref{eq:phasespacedensity}), this results in 
\begin{equation}
P=2b_{0}g(b_{0})k_{\text{B}}T_{p}/x_{0}\;\text{,}
\end{equation}
in which we have introduced $b_{0}=\sqrt{\hbar\omega x_{0}^{2}/2k_{\text{B}}T_{x}}$
and the auxiliary function $g(x)=\pi^{-1/2}\,\exp(-x^{2})/(\text{erf}(x)+1)$.
With this, the required work $W=\int_{\infty}^{0}P\,dx_{0}$ follows
as 
\begin{equation}
W=\ln2\,k_{\text{B}}T_{p}=\ln2\,k_{\text{B}}T\,\text{e}^{-2r}\;\text{.}\label{eq:Work}
\end{equation}
This result confirms the exponential decrease of $W$ with increasing
$r$, as observed in the numerical simulations. There is, however,
a certain discrepancy regarding the numerical pre-factor in the exponential
scaling, see Ref. \citep{Klaers2019} for further details. 

Since the probability density in eq. (\ref{eq:phasespacedensity})
factorizes as $\rho(\hat{x},\hat{p})=\rho(\hat{x})\,\rho(\hat{p})$
with $\rho(\hat{x})=\int_{-\infty}^{+\infty}\rho(\hat{x},\hat{p})\,d\hat{p}$
and $\rho(\hat{p})=\int_{-\infty}^{+\infty}\rho(\hat{x},\hat{p})\,d\hat{x}$,
the entropy of a squeezed thermal state results additively from the
contributions of the two quadratures: $S=S_{x}+S_{p}$. This is quite
analogous to the well known additivity of entropy for independent
subsystems. The two contributions can be determined using the Shannon
entropy, which coincides with the physical entropy in the case of
Gibbs ensembles. From $S_{p}=-k_{\text{B}}\int_{-\infty}^{+\infty}\rho(\hat{p})\,\ln(\rho(\hat{p}))\,d\hat{p}$,
one concludes that the entropy in the momentum quadrature follows
as
\begin{equation}
S_{p}/k_{\text{\text{B}}}=\ln(k_{\text{B}}T_{p}/\hbar\omega)/2+C\;\text{.}\label{eq:Sp}
\end{equation}
with an additive constant $C$. Note that this result does not reflect
the correct low-temperature behavior of the entropy, which is an artifact
of the purely classical calculation. This is, however, not crucial
for the purpose of this work. In the same way, we derive a corresponding
expression for the entropy in the position quadrature 
\begin{equation}
S_{x}/k_{\text{B}}=\ln\left[\frac{x_{0}}{b_{0}g(b_{0})}\right]-b_{0}g(b_{0})-b_{0}^{2}+C'\label{eq:Sx}
\end{equation}
During the free expansion (step i in Fig. 1) no work is performed.
The internal energy $U=\iint d\hat{x}\,d\hat{p}\,\rho(\hat{x},\hat{p})\,(\hbar\omega/2)\,(\hat{x}^{2}+\hat{p}^{2})$,
which using eq. (\ref{eq:phasespacedensity}) evaluates to 
\begin{equation}
U=\frac{k_{B}}{2}\left[(1-2b_{0}g(b_{0}))\,T_{x}+T_{p}\right]\;\text{,}\label{eq:energy}
\end{equation}
remains constant: $U(T_{x},T_{p},x_{0}\hspace{-0.3mm}=\hspace{-0.3mm}\infty)\hspace{-0.3mm}=\hspace{-0.3mm}U(T_{x},T_{p},x_{0}\hspace{-0.3mm}=\hspace{-0.3mm}0)$.
Consequently, there is no net heat flow between system and environment
and the entropy of the environment remains constant: $(\Delta S)_{\text{env}}=0$.
The total entropy change $\Delta S=(\Delta S)_{\text{env}}+(\Delta S)_{\text{sys}}$
is solely determined by the entropy change of the system $(\Delta S)_{\text{sys}}=\Delta S_{x}+\Delta S_{p}$,
which here is given by $(\Delta S)_{\text{sys}}\hspace{-0.25mm}=\hspace{-0.25mm}S_{x}(x_{0}\hspace{-0.25mm}\hspace{-0.25mm}=\hspace{-0.25mm}\hspace{-0.25mm}\infty)\hspace{-0.25mm}-\hspace{-0.25mm}S_{x}(x_{0}\hspace{-0.25mm}\hspace{-0.25mm}=\hspace{-0.25mm}\hspace{-0.25mm}0)$.
With eq. (\ref{eq:Sx}), this leads to a total entropy change of 
\begin{equation}
\Delta S=k_{\text{B}}\ln2\,\text{.}\label{eq:result_entropy}
\end{equation}
This is the expected result for an irreversible doubling of the phase
space volume. During the isothermal compression (step ii in Fig. 1)
the invested work $W$ is dissipated as heat, which leads to an entropy
increase in the environment of $(\Delta S)_{\text{env}}=W/T_{p}=k_{\text{B}}\ln2$
that exactly cancels the entropy decrease in the system $(\Delta S)_{\text{sys}}=-k_{\text{B}}\ln2$.
Thus, no net change in the total entropy occurs during this step.
The same is obviously true for the third and last step, the insertion
of the partition. This means that the total entropy change of the
universe during the erasure process solely results from the entropy
increase during the free expansion and is consequently given by eq.
(\ref{eq:result_entropy}). In total, we find that the reset of one
bit of classical information in a squeezed thermal memory leads to
the same entropy increase of $k_{\text{B}}\ln2$ as in a standard
thermal memory, while the required work can be exponentially lowered
with the squeezing factor. We expect that an experimental verification
of the predicted effect using well-established experimental platforms
such as optically trapped nano-particles \citep{Berut2012,Jun2014,Rashid2016}
and nano-mechanical devices \citep{Klaers2017} is within reach. 

Squeezed thermal environments are characterized by fast periodic amplitude
modulations in the thermal fluctuations. The significance of such
non-equilibrium thermal reservoirs stems from the fact that they may
naturally arise in systems operating in a pulse-driven fashion as
is common, for example, in digital electronics. The dissipated power
in today's microprocessors is due to both static leakage and dynamic
power dissipation, in approximately equal parts \citep{Haensch2006}.
The dynamic power dissipation in a CPU originates from the switching
of logic gates. The latter is physically realized by charging or discharging
capacitors within the gate. This process is accompanied by current
flows and associated ohmic losses. If the gate switches periodically
in time, it thus acts as a periodic heat source. As an approximation,
one can consider a gate as a point-like heat source that periodically
dissipates energy with frequency $\Omega$ in a material with heat
diffusivity $\alpha_{1}$. In this situation, fast periodic modulations
of the temperature arise that spatially extent into the environment
\citep{Michaud2005}. Such a transient temperature phenomenon is nothing
but a squeezed thermal environment, which can be seen by comparison
with eq. (\ref{eq:amplitude_modulation}). The spatial extent of this
environment can be estimated as several times the characteristic decay
length $\sqrt{\alpha_{1}/\pi\Omega}$ \citep{Michaud2005}, which
for $\Omega=1\,\text{GHz}$ and the thermal diffusivity of silicon
corresponds to several hundred nanometers. Thus, the periodic power
dissipation in a logic gate induces a squeezed thermal bath in its
surroundings that may even affect neighboring gates. The magnitude
of this effect, the squeezing factor, depends on a multitude of factors
such as geometry, thermal conductivity of materials and thermal resistance
of interfaces. Using advanced design approaches, such as thermal rectification
\citep{Roberts2011}, thermal flows can even be decoupled from electronic
currents, which further expands the possibilities to deliberately
engineer thermal environments. Similar to what has been demonstrated
in this Letter, a well-timed switching process may exploit transient
temperature phenomena to reduce the overall dissipated power. The
latter applies to all systems in which the energy costs depend on
the temperature \nobreakdash- even if they operate well above the
Landauer limit. In future, combining concepts of electronics and non-equilibrium
thermodynamics will open up new routes for more energy efficient electronics.

We thank Emre Togan, Atac Imamoglu, and Willem Vos for fruitful discussions.

\end{document}